\documentstyle[12pt]{article}
\title{\nopagebreak
\begin{flushright}
\tenrm UCTP102.98
\end{flushright}\vskip0.3in
\nopagebreak
\large \bf The Structure of AdS Black Holes \\ and \\
Chern Simons Theory in 2+1 Dimensions}
\author{Sharmanthie Fernando\thanks{email address:
fernando@physung.phy.uc.edu} and Freydoon Mansouri\thanks{email
address: Mansouri@uc.edu} \\
\it \small \it Physics Department, University of Cincinnati,
Cincinnati, OH 45221}
\date{}

\begin{document}
\maketitle

\begin{abstract}

We study 
anti-de Sitter black holes in 2+1 dimensions in terms of Chern
Simons gauge theory of the anti-de Sitter group coupled to a
source. Taking the source to be an anti-de Sitter state specified
by its Casimir invariants, we show how all the relevant features
of the black hole are accounted for. The requirement that the
source be a unitary representation leads to a discrete tower of
excited states which provide a microscopic model for the black
hole.
\end{abstract}

\section{introduction}

The BTZ solution~\cite{rone} provides a concrete and manageable
theoretical framework for testing various hypotheses concerning
classical and quantum black holes. As a result, it has been
studied extensively, as can be traced, e.g., from the review
article by Carlip~\cite{rtwo}. The main objective of this work 
is to
study how this solution can be obtained from the Chern Simons
gauge theory of the anti-de Sitter (AdS) group coupled to a
source, and the new and significant microscopic consequences
which emerge
from such a formulation. Interesting attempts linking the Chern
Simons theory to
the BTZ black hole already exist in the
literature~\cite{rthree,rfour}. However, a number of issues in
this connection need further clarification. It will be recalled
that in the BTZ formulation the black hole is a solution of
vacuum Einstein equations with a negative cosmological constant.
It differs from the standard AdS space by certain
identifications related to a discrete subgroup of the AdS group,
which changes the global topology. On the other hand, the Chern
Simons theory~\cite{rfive,rsix} is defined on a manifold $M$ with
topology $R \times \Sigma$, where $\Sigma$ is a two dimensional
space. We take the theory to be an explicit realization of the
Mach
Principle, so that in the absence of sources the field strengths
vanish and the topology is trivial (no punctures). One can then
associate non-trivial topologies to the presence of
sources~\cite{rfive,rseven,reight}. In this scenario, the
physical (metrical) space-time is the output of such a gauge
theory and should not be confused with the manifold $M$. The
physical space-time is related to a manifold $M_q$
the points $q_A$ of which are one of the canonical variables
($0+1$
dimensional fields) of the source(s)~\cite{rseven,reight}. The
presence of sources in $M$ affect not only the topology of $M$
but also the structure of $M_q$ as the
emerging
space-time. It is therefore no contradiction to state that a
Chern Simons theory in $M$ (with a source) leads to the black
hole
solution in $M_q$ (with no source).

One of the notable advantages of the Chern Simons approach is
that it allows us to express the asymptotic observables of the
theory in terms of the properties of the sources. To implement
this idea, we must identify a localized source (particle) with an
irreducible representation of the gauge symmetry
group~\cite{reight}. For the present problem, this will amount to
relating the asymptotic observables of the BTZ black hole to the
Casimir
invariants of an AdS state coupled to the Chern Simons action. We
will show that
the emerging space-time will naturally arise from such a theory
and will have all the ingredients
necessary for the AdS black hole~\cite{rnine}. These include, in
particular, the discrete subgroup underlying the identifications.
Moreover, the horizon radii of the BTZ solution are complicated
functions of the familiar AdS labels $M$ and $J$, which are
commonly referred to as ``mass'' and ``angular momentum'',
respectively. One might wonder if there is a group theoretic or
some other explanation for their functional form. We will show
that they are alternative labels for an AdS state and arise
naturally from the maximal compact subgroup of the AdS group via
induced representations.

An important consequence of the Chern Simons formulation, which
we will address in this work is the extent to
which the potential quantum aspects of the formalism will
influence the choice of the AdS representations. As mentioned
above, we take the sources which couple to the Chern Simons
action to be AdS
states, so that, to have a unitary quantum theory, these states
must be unitary representations of the AdS group. One of
the remarkable byproducts of this requirement is that the
ground state and the excited states of the black
hole form a discrete spectrum. Therefore, the Chern Simons theory
described below provides a microscopic model of the black hole
structure, which appears to be distinct from previous
suggestions~\cite{rten,releven}. 

In Section 2, we review the properties of AdS space and algebra
in a form which will be used in subsequent sections. In Section
3, we express the Chern Simons action for the AdS group in an
$SL(2,R) \times SL(2,R)$ basis. Section 4 is devoted to the
interaction with sources. Among other things, we discuss the
important role played by the constraints in relating the
invariants which label the sources to the asymptotic ovservables
of the coupled theory. In Section 5, we explore the consequences
of requiring that a source be represented by a unitary
representation of the AdS group. We will show that one of the
hitherto unexplained features of the BTZ black hole emerges from
this requirement. In Section 6, we show how the black hole space-
time emerges from the Chern Simons gauge theory described in
sections 3 through 5. In particular, we show how such features as
the periodicity of the angular coordinate and the discrete
identification group are accounted for. Section 7 is devoted to
further discussion of the results and their possible relevance to
black holes in other space-times dimensions.

\section{Anti-de Sitter space and algebra}

The anti-de Sitter space in 2+1 dimensions can be viewed  as a
subspace of a
flat 4-dimensional space with the line element
\begin{equation}
ds^2 = dX_AdX^A = dX_0^2 - dX_1^2 -dX_2^2 + dX_3^2 \end{equation}
It is determined by the constraint
\begin{equation}
(X_0)^2 - (X_1)^2 - (X_2)^2 + (X_3)^2 = l^{2}\end{equation}
where $l$ is a real constant . The set of transformations which
leave the line element invariant
form the anti-de Sitter group $SO(2,2)$. It is locally isomorphic
to $SL(2,R) \times SL(2,R)$ or $SU(1,1) \times SU(1,1)$. From
here on by
anti-de Sitter group we shall mean its universal covering group.

The AdS algebra consists of the elements $M_{AB}$ satisfying the
commutation relations
\begin{equation}
[M_{AB}, M_{CD}] = i\left(\eta_{AD} M_{BC} + \eta_{BC}M_{AD} -
\eta_{AC} M_{BD} - \eta_{BD} M_{AC}\right)\end{equation}
With $A=(a,3)$ and $a= 0,1,2,$ we can write the algebra in two
more convenient forms:
\begin{eqnarray}
M^{ab} = \epsilon^{abc}J_c  = \epsilon^{abc} 
( J^+_c +J^-_c ) \nonumber \\
M^{a3} = l\Pi^a = ( J^{+a} - J^{-a} ) \end{eqnarray}
where
\begin{equation} 
\epsilon^{012} = 1 ; \hspace{1.cm} \eta^{ab} = (1, -1,
-1)\end{equation}
Then, the commutation relations in these bases take the form,
respectively,
\begin{equation}
\left[J^a, J^b\right] = -i\epsilon^{abc}J_{c};\;\;\;\; 
\left[J^a, \Pi^b\right] = -i\epsilon^{abc}\Pi_{c};\;\;\;\;
\left[\Pi^a, \Pi^b\right] = -il^{-2}\epsilon^{abc}J_c
\end{equation} 
\begin{equation}
\left[J_a^{\pm}, J_b^+\right] = -i\epsilon_{ab}^c
J^{\pm}_{c};\;\;\;\;\;
\left[J_a^+, J_b^-\right] = 0 \end{equation}
The Casimir operators look simplest in the latter basis:
\begin{equation}
j_{\pm}^2 = \eta^{ab} J^{\pm}_a J^{\pm}_b
\end{equation}
In the other bases, they have the form,
\begin{eqnarray}
M = l^2 (\Pi^a\Pi_a + l^{-2}J^aJ_a) = 2(j_{+}^2 + j_{-}^2)
\nonumber \\
J/l = 2l\Pi_aJ^a = 2( j_+^2 - j_-^2)\;\;\;\;\;
\end{eqnarray}
We will use the same symbols for operators and their eigenvalues.

An irreducible representation of AdS group can be labeled by the
eigenvalues
of either the pair $(M,J)$ or the pair $(j_+,j_-)$. For our
applications,
it is often
advantageous to use a third set of labels which we denote by
(H,S). They correspond to 
the maximal compact subgroup $SO(2) \times SO(2)$ of $SO(2,2)$,
which is generated by $J^0$ and $\Pi^0$.
The labels $(H,S)$ are
a natural choice from the point of view of the theory of induced
representations. This can be seen from the comparison 
with the more familiar situation in the Poincar\'e group which
can be obtained from anti-de Sitter group
in the limit $l \rightarrow \infty$. From here on, we will
use the labels, $( j_+, j_- )$, $(M,J)$, and
$(H,S)$ interchangeably. The last two are related to each other
according to
\begin{equation}
M= l^2 H^2 + S^2;\;\;\;\;\;
J/l = 2lHS \end{equation}
Note that in order for $M$ to assume negative values, $H$ and $S$
must, in general, be complex.

To see the relevance of $H$ and $S$ to the BTZ solution, let us
express $H$ and $S$
in terms of the labels $M$ and $J$ by inverting Eqs. (10). We
obtain
\begin{equation}
H^2 = \frac{1}{2l^2} M \left[ 1 + \sqrt{1-
(\frac{J}{lM})^2}
\right]\nonumber\end{equation}
\begin{equation}
S^2 = \frac{1}{2} M \left[ 1 - \sqrt{1-
(\frac{J}{lM})^2}
\right] \end{equation}
For $M > 0$ and $|J| \leq lM$ ,
$H$ and $S$ are thus proportional to the horizon radii,
$r_{\pm}$, of the BTZ black hole [1]:
\begin{equation}
r_{+}/l = lH;\;\;\;\;\;\;\;\;\; r_-/l = S \end{equation}

\section{Connection and the Chern Simons action}

We begin by writing the connection in $SL(2,R) \times SL(2,R)$
basis
\begin{equation}
A_{\mu} = \omega ^{AB}_{\mu} M^{AB} = \omega ^a_{\mu} J_a +
e^a_{\mu} \Pi_a
= A^{+a}_{\mu} J_a^+ + A_{\mu}^{-a} J_a^- \end{equation}
where 
\begin{equation}
A^{\pm a}_{\mu} = \omega ^a_{\mu} \pm l^{-1} e^a_{\mu}
\end{equation}
Eqs. (14) and (15) should be viewed as definitions of $e$ and
$\omega$ in terms of $SL(2,R)$ connections. The covariant
derivative will have the form
\begin{equation}
D_{\mu} = \partial_{\mu} - iA_{\mu}= \partial_{\mu}
-iA^{+a}_{\mu} J_a^+ -i A_{\mu}^{-a} J_a^- \end{equation}
Then the components of the field strength are given by
\begin{equation}
[D_{\mu}, D_{\nu}] = -iF^{+a}_{\mu \nu} J^+_a - iF^{-a}_{\mu
\nu} J^-_a 
 = -iF^{+}_{\mu \nu}[A^+] -  iF^{-}_{\mu \nu}[A^-] \end{equation}

For  a simple or a semi-simple group, the Chern Simons action has
the form
\begin{equation}
I_{cs} = \frac{1}{4\pi}Tr \int_M A \wedge \left( dA + \frac{2}{3}
A \wedge A\right) \end{equation}
where Tr stands for trace and
\begin{equation}
A = A_{\mu} dX^{\mu} =  A^+ + A^- \end{equation} 
We require the 2+1 dimensional manifold M to have the topology 
$R\times\Sigma$, with $\Sigma$ a two-
manifold.
So, The Chern Simons action with $SL(2,R)\times SL(2,R)$ gauge
group will take the form 
\begin{equation}
I_{cs} = \frac{1}{4\pi}Tr \int_M \left[\frac{1}{a_{+}}A^+ \wedge
\left( dA^+ +
\frac{2}{3}
A^+ \wedge A^+ \right) + \frac{1}{a_{-}}A^- \wedge \left( dA^- +
\frac{2}{3}
A^- \wedge A^- \right) \right] \end{equation}
Here the quantities $a_{\pm}$ are, in general, arbitrary
coefficients, reflecting the semisimplicity of the gauge group.
Up to an overall normalization, only their ratio
is significant. It was pointed out by Witten~\cite{rfive} that in
the free Chern Simons theory the choice $a_- = -a_+$ would make
the action
proportional to Einstein's action in $M$ by imposing a metric
structure on it. Similarly, the choice $a_-
= a_+$ would give an ``exotic'' term. He also pointed out that,
in our notation, for generic values of these coefficients, the
classical equations of the free theory in $M$ remain unchanged.
In a
quantum theory~\cite{rfive}, the two terms in the action will
have a relative arbitrary coefficient.

It would be tempting to choose the first possibility on grounds
of familiarity, among other things. However, that would be an
unnatural choice from the point of view pursued here. This is
because the space-time which emerges from this theory is not the
manifold $M$ but a manifold $M_q$ corresponding to one of the
canonical variables ($0+1$ dimensional fields) of the source
which will be coupled to the Chern Simons theory in the next
section. So, the space-time is a secondary concept which emerges
from the gauge theory, and Einstein's action in $M$ plays no
direct role
in it. Moreover, in the presence of a source (or of sources), any
\'a priori choice of the coefficients $a_{\pm}$ reduces the class
of allowed holonomies, so that even the classical theory coupled
to sources will be
affected by such a choice. For these reasons, we will keep the
coefficients $a_{\pm}$ as free parameters in the sequel, so that
we can generate the correct holonomies for solutions both ouside
and inside the horizon. 

Under infinitesimal gauge transformations
\begin{equation}
u_{\pm} = \theta^{\pm\;a} J^{\pm}_{a} \end{equation}
the gauge fields transform as
\begin{equation}
\delta A_{\mu} = - \partial_{\mu} u - i[ A_\mu,u] \end{equation}
More specifically,
\begin{equation}
\delta A^{\pm\;a}_{\mu} = -\partial_{\mu}\theta^{\pm\;a} -
\epsilon^{a}_{\;bc}A^{\pm\;b}\theta^{\pm\;c}\end{equation}

As we have stated, the manifold $M$ has the topology $R \times
\Sigma$ with R representing $x^{0}$. Then subject to the
constraints
\begin{equation}
F^{\pm}_a[A^{\pm}] =\frac {1}{2}  \eta_{ab} \epsilon^{ij}
(\partial_i
A_j^{\pm\;b} - \partial_j A_i^{\pm\;b} + \epsilon^b_{\;cd}
A_i^{\pm\;c}
A_j^{\pm\;d}) = 0 \end{equation}
the Chern Simons action for $SO(2,2)$ will take the form
\begin{eqnarray}
2\pi I_{cs} = \frac{1}{a_+} \int_R dx^{0}  \int_{\Sigma}
d^2x\left(- 
\epsilon^{ij}\eta_{ab} A^{+a}_i \partial_0 A^{+b}_j +   A^{+a}_ 0
F^+_a \right)\nonumber \\
+ \frac{1}{a_-} \int_R dx^{0}  \int_{\Sigma} d^2x\left(- 
\epsilon^{ij}\eta_{ab} A^{-a}_i \partial_0 A^{-b}_j +  A^{-a}_0
F_a^- \right)\end{eqnarray}

\section{Interaction with sources}

Following the approach which has been successful in coupling
sources to Poincar\'e Chern Simons
theory~\cite{reight}, we take a source for the present problem to
be an irreducible representation
of anti-de-Sitter group characterized by Casimir invariants $M$
and $J$ (or $H$ and $S$ ). 
Within the representation, the states are further  specified by
the phase space
variables of the source $\Pi^A$ and $q^A$, $A= 0,1,2,3$, subject
to anti-de Sitter  constraints. 

For illustrative purposes, let us consider first the interaction
term for a special case  which is the analog
of the Poincar\'e case~\cite{reight} with the intrinsic spin set
to zero.
$$I_{1}  =  \int_C d\tau\left[\Pi_A D_{\tau}q^A + 
\lambda\left( q^Aq_A - l^2 \right)\right]\;\;\;\;\;\;\;\;\;\;$$
\begin{equation} 
+ \int_C d\tau\left[ \lambda_+ \left(J^{+a}J^{+}_a -l^2 j_+^2
\right) 
+ \lambda_- \left( J^{-a} J_a^- -l^2 j_-^2
\right)\right]
\end{equation}
where $C$ is a path in $M$, $\tau$ is a  parameter along $C$, and
the covariant
derivative $D_{\tau}$ is given by
\begin{equation}
D_{\tau} = \partial_{\tau} - i \omega^{AB} M_{AB} \end{equation}
The first term in this action is the same as that given in
reference~\cite{rfour}.
The second term ensures that $q^A(\tau)$ satisfy the AdS
constraint. It is not the manifold $M$ over which
the gauge theory is
defined but the space of $q's$ which give rise to the classical
space-time. The last 
two constraints identify the source being coupled to the Chern
Simons theory as an
anti-de Sitter state with invariants $j_+$ and $j_-$. These
constraints are crucial in relating
the invariants of the source to the asymptotic observable of the
coupled theory. In this respect, our action differs from that
given in reference~\cite{rfour}. Although the word
``constraints''
was mentioned there in connection with this action, they were not
explicitly stated or made use of in the sequel. 

Using the standard (orbital) representation of the generators
\begin{equation}
M_{AB} = i(q_A \partial_B - q_B \partial_A) \end{equation}
we have
\begin{equation}
\Pi_C \omega^{AB} M_{AB} q^C = \omega^{AB} ( q_A \Pi_B - q_B
\Pi_A) = \omega^{AB} L_{AB} \end{equation}
Here $L_{AB}$ are c-number quantities transforming  like
$M_{AB}$. Breaking
up this expression into $SL(2,R) \times SL(2,R)$ form just as was
done $M_{AB}$, we get
\begin{equation}
\omega^{AB} L_{AB} = A^{+a} L^+_a + A^{-a}L_a \end{equation}
So, the action $I_1$ can be written as
$$I_{1}  = \int_C d\tau\left[\Pi_A \partial_{\tau}q^A -
i\left(A^{+a}L^+_a + A^{-a}L^-_a\right)+ \lambda\left( q^Aq_A -
l^2 \right)\right]$$
\begin{equation}
+ \int_C d\tau\left[ \lambda_+ \left( J^{+a} J^{+}_a -
l^2j_+^2\right)
+ \lambda_- \left(J^{-a} J_a^{-} -
l^2j_-^{2}\right)\right]
\end{equation}
In this expression $L^{\pm}_a$ play the role of (c-number)
generalized orbital angular momenta.
iF, in addition, the representation carries generalized intrinsic
(spin) angular momenta, 
then $L^{\pm}_a$ would  have to be replaced by $J^{\pm}_a$,
respectively,
where
\begin{equation}
J^{\pm}_a = L^{\pm}_a \oplus  S^{\pm}_a 
\end{equation}

It is now clear how the interaction term $I_1$ can be generalized 
to the case when 
$S^{\pm}_a \neq 0$. We simply replace $L^{\pm}_a$ with
$J^{\pm}_a$ in $I_1$ to get
$$I_{s} = \int_C d\tau \left[\Pi_A D_{\tau}q^A + 
\lambda\left( q^A q_A - l^2 \right)\right]\;\;\;\;\;\;\;\;$$
\begin{equation}
+ \int_C d\tau \left[ \lambda_+ \left( J^{+a}
J^{+}_a - l^2j_+^2 \right)+ 
 \lambda_- \left( J^{-a} J_a^{-} 
- l^2j_-^{2}\right)\right]
\end{equation}
This expression is identical in form to that given by Eq. 26. But
now the generators are not limited to the form given by Eq. 28.
It can be expressed in a form in which the
$SL(2,R) \times SL(2,R)$ structure of the gauge group is
transparent:
$$I_{s}  = \int_C d \tau\left[\Pi_A \partial_{\tau}q^A -
(A^{+a}J^+_a + A^{-a}J^-_a)+ \lambda\left( q^Aq_A - l^2
\right)\right]$$ 
\begin{equation}
+\int_C d\tau\left[ \lambda_+ \left( J^{+a} J^{+}_a -
l^2j_+^2\right)+
\lambda_-
\left(J^{-a} J_a^- - l^2j_-^2\right)\right]
\end{equation}
In this expression $J^{\pm}_a$ play the role of c-number
generalized
 angular momenta which transform in the same way as the
corresponding generators and 
which label the source. iF there are several sources, an
interaction of the form given by Eq. 35 must be written down for
each source.

It is well known that for a Poincar\'e  state with mass $m^2 >
0$, there is a (rest)
frame in which, e.g.,
the momentum vector takes the form
\begin{equation}
p^a = (p^0, \vec{p}) \rightarrow (m,0) \end{equation}
Similarly, in the present case, there is a frame such
that when, e.g., the c-number quantity
$J^{\pm a}J^{\pm}_a > 0$, we have 
\begin{equation}
J^{\pm a} = (J^{\pm\;0}, \vec{J^\pm}) \rightarrow (j_{\pm},0)
\end{equation}
In this gauge, $SL(2,R) \times SL(2,R)$ symmetry reduces to 
$SO(2) \times SO(2)$. One can use similar methods to choose a
gauge in which the residual symmetry is, e.g., $SO(1,1) \times
SO(1,1)$.

Combining, the interaction term $I_s$ with the Chern Simons
action $I_{cs}$, we get the total action for the theory;
\begin{equation}
I = I_{cs} + I_{s} \end{equation}
In this theory, the gauge fields $A_{\mu}^{\pm}$ and the phase
space variables $q^A$ are smooth functions on the
manifold $M$. Gauge transformations on the former, which are
components of the connection in the principal $SO(2,2)$ bundle,
induce appropriate gauge transformations on the associate bundle
to which the latter belong. It is easy to check that 
the components of the field strength still vanish
everywhere except at the location of the sources.
So, the analog of Eqs. 24 becomes
\begin{equation}
\epsilon^{ij} F^{\pm\;a}_{ij} =2\pi a_{\pm} J^{\pm\;a}
\delta^2(\vec{x},\vec{x_0})
\end{equation}
In particular, when $\eta^{ab}J^{\pm}_aJ^{\pm}_b > 0$, we get, in
the special (rest) frame
\begin{equation}
\epsilon^{ij} F^{\pm\;0}_{ij} = 2\pi a_{\pm} j_{\pm}
\delta^2(\vec{x},\vec{x^0})
\end{equation}
All other components of the field strength vanish. We thus see
that because of the 
constraints appearing in the action given by Eq. 34, the strength
of the
sources corresponding to the maximal compact subgroup of the
gauge group
become identified
with their Casimir invariants. These invariants,
in turn, determine the asymptotic observables of the theory.
Since such 
observables must be gauge invariant, they are expressible in
terms of Wilson loops,
and a Wilson loop about our source can only depend on, e.g.,
$j_+$ and $j_-$.

From the data on the manifold $M$ given above, it is possible to
determine the properties of the emerging space-time. To this end,
we note that in the gauge in which Eq. 39 holds, the only
non-vanishing components of the gauge potential are given by
\begin{equation}
A^{\pm 0}_{\theta} = 2a_{\pm} j_{\pm}
\end{equation}
where $\theta$ is an angular variable. As an example, consider
the case of $a_+ = a_- = 1$. Then, using Eqs. 14 and 15, the
non-vanishing components can also be written as,
\begin{equation}
e^a_{\theta}/l = (j_+ - j_-) = r_-/l
\end{equation}
\begin{equation}
\omega^a_{\theta} = (j_+ + j_-) = r_+/l
\end{equation}
Although these are components of a connection which is a pure
gauge, they give rise to non-trivial holonomies
around the source. More explicitly, we have
\begin{equation}
W[e] = P exp^{ \int_{\gamma} e^{0}_{\theta} \Pi_0}
\end{equation}
\begin{equation}
W[\omega] = P exp^{ \int_{\gamma} \omega^{0}_{\theta}J_0}
\end{equation}
Here, $\gamma$ is a loop around the source, which can be
represented as a map from the circle to 
the manifold $M$, i.e., 
$ \gamma : S^1 \rightarrow M$ with
$\gamma( \sigma + 2\pi) = \gamma(\sigma)$. These holonomies are
not gauge invariant~\cite{rtwo} and transform by conjugation
under $SO(2)\times SO(2)$ transformations.

The holonomies $W[e]$ and $W[\omega]$ control the parallel
transport of a
vector such as $q^A$ around the loop $\gamma$ in $M$.
Since $\gamma$  is non trivial, the initial and the final vector
will differ from each
other by a factor involving $W[e]$ and $W[\omega]$. Since the
quantities $e_{\phi}^{0}$ and 
$\omega_{\phi}^{0}$ are components of a
``Flat'' connection, the holonomy can only depend on
the homotopy class of 
the loop $\gamma$. As a result, the quantities $W[e]$ and
$W[\omega]$  
generate the fundamental group $\Gamma$ of the manifold $M$ in
the presence of a source.
Since $\gamma(\sigma)$ is periodic, $\Gamma$
becomes
a discrete subgroup of $SL(2,R) \times SL(2,R)$.

\section{Restriction to unitary representations}
We have indicated that our sources transform as irreducible
representations of the AdS group. From purely classical
considerations, the choice between unitary and non-unitary
representations might not seem to be relevant. But to allow for
the possibility of quantizing the Chern Simons theory
consistently, we will require that our sources be represented by
unitary representations of AdS group. As we shall see, this
requirement will also have interesting consequences for the
classical space-time which emerges from this theory.

Since the AdS group in $2+1$ dimensions can be represented in the
$SL(2,R) \times SL(2,R)$ form, we can construct the unitary
representations of $SO(2,2)$ from those of $SL(2,R)$. We will
assume that the reader is familiar with the representation theory
of $SL(2,R)$. Here, we will state a few facts relevant to its
unitary representatons~\cite{rtwelve}. More information about
these representations can be found in papers listed in
reference~\cite{rtwelve} and those cited therein. The states in
an irreducible representation of $SL(2,R)$ are specified by the
eigenvalues of its Casimir operator $j^2$ (see Eq. 8) and, e.g.,
the element $J_0$, where we have suppressed the super-
(sub)scripts $\pm$ distinguishing our two $SL(2,R)$ 's. Thus, we
have
$$j^2 |\Phi, F, m> = \Phi (\Phi + 1) |\Phi, F, m>$$
$$J_0 |\Phi, F, m> = (F + m) |\Phi, F, m>$$
In these expressions, $\Phi$ is, in general, a complex number, F
is a fraction, and m is an integer. It is well
known that $SL(2,R)$ has four series of unitary
representations~\cite{rtwelve}, all of which are infinite
dimensional. For the present application, we choose the discrete
series in which each irreducible representation is an infinite
tower of states for which the eigenvalues of $J_0$ are bounded
from below. That is,
\begin{equation}
F = - \Phi = real \;\; non-integer \;\; number > 0; \;\;\;\;\; 
m = 0, 1,
2,...
\end{equation}
So, for this series,
in the notation of section 2, the eigenvalues of the Casimir
invariants of $SL(2,R) \times SL(2,R)$
can be written as, 
\begin{equation}
j^2_{\pm} = F^2_{\pm} - F_{\pm}
\end{equation}
It follows that the infinite set of states can, in a somewhat
redundent notation, be
specified as
\begin{equation}
 |j_{\pm}^2, F_{\pm}+ m_{\pm}>; \;\;\;\;\; m_{\pm} = 0, 1, 2, ...
\end{equation}
Clearly, the integers $m_{\pm}$ are not necessarily equal.
Using these states, we can construct the discrete series of the
unitary
representations of $SO(2,2)$. A typical state
will have the following labels:
\begin{equation}
| M, J > = | j_+^2, j_-^2, F_+ + m_+, F_- + m_- >
\end{equation}

As a prelude to identifying the labels $M$ and $J$ with the
corresponding labels in the BTZ solution in the next section, let
us consider the physical restrictions imposed on $F_{\pm}$. It is
clear from Eqs. 11 through 13 that, up to proportionality
constants, we want to identify $r_{\pm}^2$ with $H^2$ and $S^2$,
respectively. To do so, we must choose positive square roots of
$H^2$ and $S^2$ since the radii are intrinsically positive
quantities. We also see from these equations that to have real
non-zero horizon radii, we must have $M > 0$. Then, since the two
$SL(2,R)$ 's appear symmetrically in the formalism, we must take
$j_{\pm}^2$ to be real and positive. Then, Eq. 46 requires that
$F_{\pm} > 1$. Once this condition is satisfied, it can be seen
from Eq. 9 that $|J/l| \leq |M|$, as required in the
BTZ formalism. The extreme case corresponds to $j_-^2 = 0$. This,
in turn, requires that $F_- = 1$. We note that similar statements
also hold
for the Euclidean version of the AdS
space, where the symmetry group becomes $SO(1,3)$.

So far we have discussed the unitary discrete series of
$SL(2,R)$, and
subsequently of $SO(2,2)$, for which the eigenvalues of
$J_0^{\pm}$ are bounded from
below. We have seen that they are suitable candidates for the
microstates of the AdS black hole. Of the other three
series of unitary representations of $SL(2,R)$, the principal
series are characterized by eigenvalues of the Casimir operator
which are negative. Using the results of section 2, it is easy to
show that they lead to negative values of $M$ and to complex
values of the horizon radii. On this basis alone, we consider
them not relevant to the description of the black holes. The
other two
series are the supplementary series and the discrete series for
which $\Phi$ and $F$ are equal and negative. They cannot be ruled
out on the basis of the criteria discussed above. However, for
these series, as well as for the principal series, the
eigenvalues of the operators $J_0^{\pm}$ are not bounded from
below. In analogy with real $3+1$ dimensional world, if we
identify the corresponding microstates as physical degrees of
freedom, then we will have to identify $J_0^{\pm}$ with some
physical observables. It is difficult to conceive of a physical
observable with infinitely large negative eigenvalues. Until such
an observable could be justified on physical grounds,
we seem to be limited to the first discrete series discussed
above.

\section{The black hole space-time}

To see how the space-time structure emerges from our anti-de
Sitter gauge theory, we follow an approach which led to
the emergence of space-time from
Poincar\'e~\cite{reight} and super Poincar\'e~\cite{rthirteen}
Chern Simons gauge
theories. We have emphasized that the manifold $M$ is not to be
identified with space-time. But the information encoded in it and
discussed in section 4, is sufficient to fix the properties of
the emerging space-time. To this end, let us consider a manifold
$\hat{M}_q$ satisfying the AdS constraint
\begin{equation}
\hat{q}_0^2 - \hat{q}_{1}^2 - \hat{q}_2^2 + \hat{q}_3 ^2 = l^2 =
-\Lambda^{-1}
\end{equation}
where $\Lambda$ = cosmological constant. In fact, our $SL(2,R)
\times SL(2,R)$ formulation allows us to take $\hat{M}_q$
to be the universal covering
space of the AdS space. As we shall see, the emerging space-time
is the quotient of $\hat{M}_q$ by the discrete subgroup
$\Gamma$
discussed in section 4. Moreover, the source coupled to the Chern
Simons action is an AdS state characterized
by the Casimir invariants $(M,J)$ or, equivalently, $(H,S)$. To
parametrize $\hat{M}_q$ consistent
with the above constraint, consider a pair of 2-vectors,
\begin{equation}
\vec{\hat{q}}_{\phi} =  (\hat{q}^1, \hat{q}^2) = ( f cos {\phi},
f sin {\phi})
\nonumber \end{equation}
\begin{equation}
\vec{\hat{q}}_t = ( \hat{q}^0, \hat{q}^3) = \left( \sqrt{f^2 +
l^2} cos(t/l),
\sqrt{f^2 + l^2} sin(t/l) \right) 
\end{equation}
where $ f= f(r)$, with $r$ a radial coordinate which for an
appropriate $f(r)$ will
become the radial coordinate appearing in the line element for
the BTZ black hole. As far the
constraint given by Eq. 49 is concerned, the functional form of
$f(r)$ is
irrelevant. The parameters $\phi$ and $t/l$ are
both periodic. We will keep $\phi$ periodic throughout. However,
since we are taking $\hat{M}_q$ to be universal
covering
space of AdS space, we do not have to , and we will not, identify
$t$ with $t + 2\pi l$. With or without this identification, the
vectors $q^A$ parametrized in this fashion do not behave in the
same way as the vectors in the manifold $M$ when they are
parallel transported along a loop encircling the source.
Computing the line element in terms of the parameters
$(t/l,r,\phi)$, we get
\begin{equation}
ds^2 = (1 + \frac{f^2}{l^2}) dt^2 -
\frac{f^{'\;2} dr^2}{(1 + \frac{f^2}{l^2})} 
- f^2 d\phi^2 
\end{equation}
where ``prime'' indicates differentiation with respect to $r$.

Anticipating the results to be given below, let us 
compare this line element with that for the
BTZ black hole~\cite{rone}.
\begin{equation}
ds^2= [\frac{r^2}{l^2} -M + \frac{J^2}{4r^2}] dt^2 -
\frac{dr^2}{[\frac{r^2}{l^2} -M + \frac{J^2}{4r^2}]} - r^2 [d\phi
- \frac{J^2}{2r^2}dt]^2 
\end{equation}
If we identify the labels $M$ and $J$ with the Casimir invariants
of an irreducible representation of the AdS group as discussed in
the previous sections,
we see that the line element given by Eq. 52 corresponds to an
irreducible representation with $J= 0$ and $M = -1$. Such a state
will not correspond to any of the series of the unitary
representations of the AdS group discussed in the previous
section. Moreover, as
we have noted in connection with Eqs. 11 and 12, for these
values
of $J$ and $M$, the invariant $H$ is pure imaginary. This,
in turn, implies that the quantities $r_{\pm}$ will also be
imaginary. Thus, we can interpret the line 
element in Eq. 52 as a special form of the BTZ line element which
has
been ``Wick rotated'' into the imaginary axis in the
complex $H$ space. In this form, the consequences of the residual
gauge
transformations involving $H$ and $S$, or $r_{\pm}$, which we
will perform
below on $\hat{q}^A(\tau)$ become very similar to those performed
in
the
Poincar\'e~\cite{rseven,reight}
Chern Simons gravity. We must keep in mind, however, that in
the end, we must Wick rotate the results back to the real
$r_{\pm}$ axes so that the source coupled to the Chern Simons
theory would belong to a unitary representation and that the
resulting horizon radii would be real. We thus see that the
choice of a unitary representation has interesting classical
consequences.

With these issues in mind, we want to obtain the space-time
manifold $M_q$ by performing appropriate gauge
transformations on $\hat{M}_q$. Although the original theory was
invariant under  $SL(2,R) \times SL(2,R)$ gauge transformations,
we have already
reduced this symmetry by choosing to work in a gauge in which
Eq. 39 holds.
In fact, the left over symmetry is just $SO(2) \times SO(2)$
generated, respectively, by $J_0$ and
$\Pi_0$, or, equivalently, by $J^{\pm\;0}$. So, identifying the
parameters $\phi$ and
$t/l$, respectively, with each $SO(2)$, 
consider the local gauge transformation
\begin{equation}
\vec{\hat{q}'}_{\phi}(\phi) = e^{i\frac{r_+}{l} \phi J^0}
\vec{\hat{q}}_{\phi} ( \phi)\end{equation}
It leaves $\vec{\hat{q}}_t(t/l)$ invariant. Then, since $\phi$ is
$2\pi$ periodic,
\begin{equation}
\vec{\hat{q}}_{\phi'}(\phi + 2 \pi) = e^{i 2 \pi \frac{r_+}{l} 
J^0} \vec{\hat{q}}_{\phi} ( \phi) \end{equation}
Similarly, consider the gauge transformation
\begin{equation}
\vec{\hat{q}'}_{t}(t/l,\phi) = e^{ir_- \phi \Pi^0}
\vec{\hat{q}}_{t}
(t/l) \end{equation}
It leaves $\vec{\hat{q}}_{\phi}$ invariant and leads to
\begin{equation}
\vec{\hat{q}}_{t'}(t/l, \phi + 2 \pi) = e^{ir_-2\pi \Pi^0}
\vec{\hat{q}}_{t} (t/l,\phi) \end{equation}
We note that the factors picked up by $\vec{\hat{q}}_{\phi}$ and
$\vec{\hat{q}}_{t}$ under rotation by $2\pi$ are the same as
those given by the holonomies given by Eqs. 43 and 44.
Thus, the periodicity of $\phi$ has led to a discrete subgroup
of isometries in the universal covering space of the AdS space.
The parameters $\frac{ 2 \pi}{l} r_{\pm}$ for these
transformations were 
chosen to demonstrate the connection between the holonomies in
$M$ and the 
identifications necessary for the BTZ black hole. Moreover, in
contrast to the situation for the Poincar\'e
group,
the residual symmetry $SO(2) \times SO(2)$ assigns symmetrical
roles to the invariants $(H,S)$ or $(r_+, r_-)$ as well as the
parameters $\phi$ and $t/l$.
To reflect this symmetrical role, we can perform our gauge
transformations on $\vec{\hat{q}}_{\phi}$ and
$\vec{\hat{q}}_{t}$ in the following more symmetrical manner:
\begin{eqnarray}
\vec{q}_{\phi'}(\phi, t/l) = e^{i \left( \frac{r_+}{l}
\phi - \frac{r_-t}{l^2} \right)J^0} \vec{\hat{q}}_{\phi} ( \phi)
\nonumber \\
\vec{q}_{t'} (t/l,\phi) = e^{i \left( \frac{r_-}{l} \phi
- \frac{r_+t}{l^2} \right)l\Pi^0} \vec{\hat{q}}_{t} (t/l)
\end{eqnarray}
It then follows that
\begin{eqnarray}
\vec{q}_{\phi '}(\phi + 2 \pi, t/l) = e^{i 2
\pi\frac{r_+}{l} J^0} \vec{q}_{ \phi '} ( \phi, t/l)
\;\;\;\;\;\;\;\;\;\;\nonumber \\
\vec{q}_{t '}(t/l,\phi + 2 \pi) = e^{i 2
\pi\frac{r_-}{l} l\Pi^0} \vec{q}_{t} (t/l, \phi)
\;\;\;\;\;\;\;\;\;\;\;\nonumber \\
\vec{q}_{\phi '}(\phi + 2 \pi, t/l + 2 \pi) = e^{i 2 \pi
\left( \frac{r_+}{l}  - \frac{r_-}{l} \right) J^0}
\vec{q}_{\phi'} ( \phi, t/l)\nonumber \\
\vec{q}_{t'}(t/l +2\pi,\phi + 2 \pi) = e^{i 2 \pi
\left(\frac{r_-}{l} - \frac{r _+}{l} \right)l\Pi^0}
\vec{q}_{t'} ( \phi, t/l) \end{eqnarray}
Thus, given the previous identifications, the last two
expressions do not lead to any new identifications. We can now
write
\begin{equation}
\vec{q}_{\phi '} ( \phi, t/l) = \vec{q}_{\phi '} (
\phi '); \hspace{1.0cm}
\vec{q}_{t '} ( \phi, t/l) = \vec{q}_{t'} ( t '/l)
\end{equation}
where 
\begin{equation}
\phi ' = \frac{ r_+}{l} \phi  -  \frac{r_-t}{l^2};
\hspace{1.0cm}
 t ' = \frac{ r_-}{l} \phi -  \frac{r_+t}{l^2} \end{equation}
Now we note again that the vector $(q_{\phi'}, q_{t'})$
transforms in the same way as the one which in section 4 was
parallel
transported around a loop in the manifold $M$. Calling the
manifold to which such vectors belong $M_q$, we see that
this
manifold incorporates the same dynamics as the phase space
variables in
$M$, and we are justified in using the same letter $q$ for both.
Thus, we can parametrize the manifold $M_{q}$ as
follows:
\begin{eqnarray}
 q^1 = f cos \left( \frac{r_+}{l}\phi - \frac{r_-t}{l^2}
 \right)\;\;\;\;\;\;\;\;\nonumber \\
 q^2 = f sin \left( \frac{r_+}{l}\phi - \frac{r_-t}{l^2}
\right)\;\;\;\;\;\;\;\;\nonumber \\
q^0 = \sqrt{ f^2 + l^2 } cos \left( \frac{r_-}{l}\phi -
\frac{r_+t}{l^2} \right)\nonumber \\
q^3 = \sqrt{ f^2 + l^2 } sin  \left( \frac{r_-}{l}\phi -
\frac{r_+t}{l^2} \right) \end{eqnarray}
From these we can compute the line element. It is given by
\begin{equation}
ds^2 =  \frac{f^2}{l^2}\left(r_+ d\phi - r_-\frac{dt}{l}\right)^2
- \frac{ f^{'\;2} dr^2}{(1 + \frac{f^2}{l^2}
)} -\left(\frac{f^2}{l^2}-1\right) \left( r_- d\phi - r_+
\frac{dt}{l}
\right)^2 \end{equation}

It will now be recalled that the quantities $r_{\pm}$ appearing
in this expression are ``Wick rotated'' relative 
to the  corresponding invariants which appear in the 
BTZ solution. We must, therefore, rotate  them back to the Re
$r_{\pm}$ axes by letting
\begin{equation}
r_{\pm} \rightarrow -i r_{\pm} \end{equation}
Then, we get
\begin{equation}
ds^2 =  -\frac{f^2}{l^2}\left( r_+ d \phi -
r_-
\frac{dt}{l}\right)^2 - \frac{ f{'\;2} dr^2}{(\frac{f^2}{l^2} +
1)} + 
\left(\frac{f^2}{l^2}-1\right) \left( r_- d\phi - r_+
\frac{dt}{l} \right)^2
\end{equation}
Finally, to put this expression in a form identical to that given
by BTZ~\cite{rone} given by Eq. 53, we set
\begin{equation}
 \frac{f^2}{l^2} =  \frac{ r_-^2 - r^2}{r_+^2 - r_-^2};
\hspace{0.5in} r < r_- \end{equation}

It can be seen from Eqs. 61 and 66 that the parametrization
leading to this
expression is valid for $r < r^-$ and any value of the
parameter $l$. The
simplest way of obtaining suitable parametrizations for all
values of $l$ is to observe that parametrization in terms of
circular functions are Wick rotated relative to the BTZ solution.
Then, as can be seen from Eq. 64, when we rotate the Casimir
invariants $r_{\pm}$ back to their real axes in their respective
complex $r_{\pm}$ planes, as we did in the above example, we are
effectively replacing trigonometric functions by their
corresponding hyperbolic functions. We emphasize that this
replacement leaves the periodicity of the angle $\phi$ intact
since the Wick rotation occurs not in $\phi$ but in complex $r_+$
and $r_-$ spaces. This means that we do not need to impose
periodicity on $\phi$ ``by hand'' if we wish to use a hyperbolic
parametrization~\cite{rone,rtwo} which is advantageous in many
instances.

It is, nevertheless, of interest to see if a parametrization in
terms of circular functions works for $r>r^+$. For this to be
possible, there must be a class of holonomies in $M$ which are
consistent with such a parametrization. Consider the
following expressions:
\begin{eqnarray}
q^1 = f cos \left( \frac{r_-}{l}\phi - \frac{r_+t}{l^2}
\right)\;\;\;\;\;\;\nonumber \\
q^2 = f sin \left( \frac{r_-}{l}\phi - \frac{r_+t}{l^2}
\right)\;\;\;\;\;\;\nonumber \\
q^0 = \sqrt{ f^2 + l^2 } cos \left( \frac{r_+}{l}\phi -
\frac{r_-t}{l^2} \right)\nonumber \\
q^3 = \sqrt{ f^2 + l^2 } sin  \left( \frac{r_+}{l}\phi -
\frac{r_-t}{l^2} \right) \end{eqnarray}
This parametrization of $M_q$ corresponds to the class of
holonomies in $M$ for which the parameters $a_{\pm}$ in Eqs. 20
and 40
are given by $a_+ = -a_- =1$. 
Then, we can get back the BTZ metric of Eq. 53 by computing the
line
element in terms of these parameters, using Eq. 64 for inverse
Wick
rotation, and setting $f$ to
\begin{equation}
 \frac{f^2}{l^2} =  \frac{ r^2 - r_+^2}{r_+^2 - r_-^2};
\hspace{0.5in} r > r_+ \end{equation}

\section{The microscopic black hole structure}

The formalism discussed in the previous sections provides a
framework for introducing internal structure for black holes. In
the metrical approach used by BTZ~\cite{rone} to obtain the AdS
black hole, all one can infer is that the black hole is endowed
with the two Casimir invariants $M$ and $J$ of the asymptotic AdS
group.
There is no room in this approach for the introduction of an
internal structure for the black hole. In our approach in which
we take a source to be a state belonging to an irreducible
representation of the AdS group, an internal structure for the
black hole arises naturally. When such a source is coupled to the
Chern Simons action, the emerging black hole is still labeled by
the two Casimir invariants $M$ and $J$ or, equivalently, $H$ and
$S$. However, for a given $M$ and $J$, the irreducible
representation is a Hilbert space the states of which may be
viewed as the internal states of the black hole. For unitary
representations, the Hilbert space is infinite dimensional. The
explicit representations which were discussed in detail in
section 5 were the discrete series bounded from below. Each
representation is
determined by a ``ground state'' labeled by quantities $F_+ > 1$
and $F_- > 1$, which determine the two Casimir invariants of the
AdS group according to Eq. 46 and, consequently, the horizon
radii and the area associated with the black hole. For each
ground state, there is an
infinite tower of states with labels which differ from those of
the ground state by two separate integers. So, the black hole
acquires the degrees of freedom which would be absent in a
standard general relativity approach.

The level structure exhibited in this model is reminiscent of the
bound state structure familiar from atomic physics except that
the
the energy of the ground state is positive. In this respect, we
note that in the BTZ solution and the subsequent works the labels
$M$ and $J$ have been identified as ``mass'' and ``angular
momentum''.
On the other hand, from the point of view of induced
representations, it is the labels arising from the maximal
compact subgroup, in this case $SO(2) \times SO(2)$, which are
more suitable for such designations. In other words, it is the
eigenvalues of $\Pi^0$ and $J^0$ which we identify as
``energy'' and ``spin'', respectively. 

The general features of the formalism developed in this work in
$2+1$ dimensions are applicable to black holes in any dimension.
A typical black hole is specified in terms of its asymptotic
observables. If we identify these observables with the Casimir
invariants of the asymptotic symmetry group, usually a noncompact
group, then the corresponding Hilbert space could serve as a
microscopic model for the black hole. It remains to be seen
whether
such models, and their modifications to take into account the fact that the 
symmetry involved here is a local gauge symmetry and not just a 
a global one, are sufficiently realistic.

\bigskip
We would like to thank F. Ardalan for a reading of the manuscript
and for many helpful comments. This work was supported, in part
by the Department of Energy
under the contract number DOE-FGO2-84ER40153.

\end{document}